# Evaluating hybrid controls methodology in early-phase oncology trials: a simulation study based on the MORPHEUS-UC trial

Short title

**Hybrid control methods in early-phase trials: a simulation study**

Guanbo Wang[1,2,3], Melanie Poulin Costello[3], Herbert Pang[4], Jiawen Zhu[4], Hans-Joachim Helms[5], Irmarie Reyes-Rivera[5], Robert W. Platt[6,7], Menglan Pang[8], Artemis Koukounari[9]

1. CAUSALab, Harvard T. H. Chan School of Public Health, Boston, USA
2. Department of Epidemiology, Harvard T. H. Chan School of Public Health, Boston, USA
3. F. Hoffmann-La Roche Ltd, Mississauga, ON, Canada
4. Genentech, South San Francisco, CA, USA
5. F. Hoffmann-La Roche Ltd, Basel, Switzerland
6. Department of Epidemiology, Biostatistics and Occupational Health, McGill University, Montreal, QC, Canada
7. Department of Pediatrics, McGill University, Montreal, QC, Canada
8. Biogen, Cambridge, MA, USA
9. F. Hoffmann-La Roche Ltd, Welwyn Garden City, UK

**Corresponding author**
Name: Melanie Poulin-Costello
Address: 7070 Mississauga Rd Mississauga ON L5N 5M8
Phone: 647-202-0152
Email: Melanie.poulin-costello@roche.com

**Abstract**

Phase Ib/II oncology trials, despite their small sample sizes, aim to provide information for optimal internal company decision-making concerning novel drug development. Hybrid controls (a combination of the current control arm and controls from one or more sources of historical trial data [HTD]) can be used to increase the statistical precision. Here we assess combining two sources of Roche HTD to construct a hybrid control in targeted therapy for decision-making via an extensive simulation study. Our simulations are based on the real data of one of the experimental arms and the control arm of the MORPHEUS-UC Phase Ib/II study and two Roche HTD for atezolizumab monotherapy. We consider potential complications such as model misspecification, unmeasured confounding, different sample sizes of current treatment groups, and heterogeneity among the three trials. We evaluate two frequentist methods (with both Cox and Weibull accelerated failure time [AFT] models) and three different commensurate priors in Bayesian dynamic borrowing (with a Weibull AFT model), and modifications within each of those, when estimating the effect of treatment on survival outcomes and measures of effect such as marginal hazard ratios. We assess the performance of these methods in different settings and potential of generalizations to supplement decisions in early-phase oncology trials. The results show that the proposed joint frequentist methods and noninformative priors within Bayesian dynamic borrowing with no adjustment on covariates are preferred, especially when treatment effects across the three trials are heterogeneous. For generalization of hybrid control methods in such settings we recommend more simulation studies.

# 1. INTRODUCTION

The traditional drug development series of clinical trial phases need robust tools for decision-making to advance a drug beyond the early phases where trials have small sample sizes and may be further subset by targeted therapy and their biomarker subgroups. Phase Ib/II trials inform decisions for furthering development of novel treatments, ultimately for a confirmatory, randomized clinical trial (Phase III). Most oncology Phase Ib/II trials are designed to explore safety and preliminary efficacy signals as proof of concept and are therefore usually not fully powered to detect a certain treatment effect.

Estimates of potential treatment effects in these smaller trials with limited sample sizes lack statistical precision, thereby compromising optimal decision-making for advancing a drug in development. In cases where limited numbers of patients are used in Phase Ib/II trials, we are motivated to explore the use of historical trial data (HTD) as a hybrid control (a combination of the current control arm and HTD control arm[s]) in a nonconfirmatory context where the future development of an investigational treatment is of interest.

Incorporating data external to an ongoing clinical trial is important to supplement smaller sample sizes where informed decisions for advancing a drug to the next phase of development are needed. We define external data as any source of clinical data generated outside of a current Phase Ib/II clinical trial inclusive of HTD and real-world data from routine electronic health record data, insurance claims data, or patient registries. The use of HTD has many advantages over real-world data sources in that the data were collected as part of a clinical trial protocol; therefore, unlike real-world data, they are of registrational intent and quality, have protocol specified endpoints collected at regular time intervals (such as CT scans), and apply guidance and conventions for clinical trials (such as RECIST). Additionally, HTD can be more consistent with Pocock's criteria published in 1976,[1] in terms of their implementation in the same organization and same requirements for patient eligibility, as well as precisely defined standard treatment and similarity of treatment and of treatment evaluation with the randomized control. High-quality patient-level HTD are readily available within a company and can also be shared among companies via mechanisms such as the TransCelerate HTD Sharing initiative (https://www.transceleratebiopharmainc.com/initiatives/historical-trial-data-sharing/).

The current evolution in the development of associated statistical methodologies has attracted interest in the implementation of hybrid controls.[2-4] A hybrid control arm consists of supplementing the control arm from the current trial with patient-level data from the HTD control.[2] The use of external controls has been explored in confirmatory trials with larger sample sizes[2-6] and in the context of only one external data source, or three or more sources based on individual patient data meta-analysis.[7-10] However, we assess combining two sources of HTD to construct a hybrid control for the current control arm of a Phase Ib/II study in targeted therapy for decision-making in oncology. HTD is sourced from trials conducted to study another investigational drug. In the next paragraph we describe these Roche data sources (i.e., Phase Ib/II study and HTD). It should also be noted that in all these data, there might be different levels of heterogeneity in terms of the study population among the current trial and historical ones as well as between historical trials.

Roche has launched the MORPHEUS[11] Phase Ib/II platform of umbrella studies designed to identify early signals of safety and anticancer activity. One of these studies assesses treatments for urothelial carcinoma (UC) (NCT03869190). MORPHEUS-UC randomizes patients to one common control arm of the cancer immunotherapy atezolizumab (atezo mono) or one of eight experimental arms with targeted therapy in combination with atezolizumab (atezo combo). Patients remain on treatment until disease progression death. Enrollment into the experimental arms takes place in two stages: a preliminary stage, followed by an expansion stage. Approximately 15 patients are enrolled during the preliminary stage in most of the treatment arms. If clinical activity is observed in an experimental arm during the preliminary stage, approximately 25 additional patients may be enrolled in that arm during the expansion stage.

HTD for atezo mono in UC is available from two Roche studies: IMvigor 210[12] (NCT02951767), an ongoing Phase II single-arm study of atezo mono, and the experimental atezo mono arm of the Phase III randomized study IMvigor 211[13] (NCT02302807).

As the performance of different statistical methods for hybrid controls can vary in different settings, we use simulation to obtain empirical results. Simulation studies can assess the violation of assumptions of statistical methods and quantify the robustness of methods through bias and error measurements.[14] Simulation approaches can improve our understanding of the performance of alternative statistical methods for hybrid controls. The use of hybrid controls has also been encouraged by global regulatory agencies.[15-18]

Our simulation study is based on the data of one of the experimental arms and the atezo mono of MORPHEUS-UC as well as the two Roche HTD for atezo mono. The objective of this paper is to evaluate both existing and newly extended methods with different models and specifications in estimating the treatment effect. We aim to make assessments and explore the potential of generalizations to support early-phase oncology trials in the following settings: 1) borrowing data from two historical trials, and 2) survival outcome. Potential complications involve the presence of the following: 1) model misspecification, 2) unmeasured key baseline covariates that might be related to outcome and treatment groups in the current trial and HTD (unmeasured confounding), 3) using different sample sizes for the treatment arm, 4) different magnitudes of treatment effect among the three data sources, and 5) different levels of heterogeneity among the three trials.

The remainder of the paper is organized as follows: In Section 2 (Methods), we describe analytically the examined statistical methods for hybrid controls that merit exploration for our setting. Next, in Section 3 (Simulation), we lay out the simulation details, followed by the results in Section 4 and discussion in Section 5.

## 2. METHODS

We evaluate two frequentist approaches and three different commensurate priors in Bayesian dynamic borrowing,[19] with two different modifications within each of those, to estimate the parameter of interest, average treatment effect (ATE) among the subjects in the current (MORPHEUS-UC) trial. More precisely, the ATE in our simulation study is the progression-free survival hazard ratio (HR) at a population level of the same patients under different treatment conditions being compared for each treatment (atezo combo versus atezo mono). We restrict ourselves to the ATE in this work, but the heterogeneous treatment effect (HR conditional on the potential effect modifiers) can be explored in future work. In this section, we review and introduce in turn each of the considered methods and then make a comparison among them. To apply all these methods, since we only have the outcomes corresponding only to the treatment received for each patient, we need certain assumptions, which we briefly summarize here. The first assumption is positivity, which requires that patients have a nonzero probability of receiving the control treatment (atezo mono) via enrollment into the current trial or HTD; the second is consistency, which requires that there are no different forms or versions of each treatment level that lead to different potential outcomes. In particular, the version of the control treatment (atezo mono) is the same across the current trial and HTD data sources; and the third is exchangeability, which requires, conditional on the observed confounders, the potential outcomes and the treatment assignment are independent; similarly, conditional on the same set of observed confounders, the potential outcomes and the data sources are also independent.[20] Note that we aim to apply each of these methods in the following setting: the sample sizes of the current MORPHEUS-UC experimental treatment arm (denoted by MT) and of the current MORPHEUS-UC control arm (denoted by MC) are both small, while the sample sizes of the two HTD that contain eligible historical controls (denoted by HC0 and HC1) are both relatively large.

**2.1 Frequentist approaches**

In general, frequentist approaches to the use of HTD aim to balance the population of current and HTD treatment arms in terms of the probability of treatment assignment conditional on measured baseline covariates (i.e., propensity score) for a less biased comparison. In this work, we evaluate two propensity score methods: inverse probability of treatment weighting (IPTW)[21-24] and matching.[25-28] IPTW directly employs a function of propensity scores as the weights of the individual subjects, while generally the weights in matching do not equal the (estimated) propensity scores. The matching method used here, full matching (FM)[29-30] (also referred to as optimal full matching) assigns every subject randomized to the experimental treatment arm and the control subjects in the HTD to a single subclass each. The advantage of FM is that it minimizes the mean absolute within-subclass distances in the matched data by the chosen number of subclasses and the assignment of subjects to subclasses. Weights are then computed based on subclass membership[30] so that the propensity scores are used to create the subclasses but not to form the weights directly. As a consequence, FM is considered less sensitive to the parametric assumptions of the propensity score model than is IPTW.[31-34]

In both the evaluated frequentist propensity score methods, the hazard of the occurrence of the event of interest is regressed on an indicator variable denoting the treatment assignment using a Cox proportional hazards model that incorporates the appropriate set of weights, and the estimate of the parameter of interest (also referred to as causal estimand or causal quantity) is the coefficient of the treatment from this model. By causal estimand or causal quantity we mean a mathematically specified expression within specific statistical methods, which we distinguish from the "estimand" defined in the addendum to the ICH E9(R1) guideline 2019 as recommended by Ho et al. 2020.[35] Because the causal estimand is the ATE among the subjects in the current (MORPHEUS-UC) trial, we briefly explain how the two frequentist approaches adjust the HTD to resemble the subjects in the current (MORPHEUS-UC) trial. For instance, the weight of each subject in MC and MT in the current trial is 1 (leaving the MORPHEUS-UC subjects untouched). The weight for the subjects from HTD is $\frac{ps}{1-ps}$, where $ps$ is the estimated propensity score in IPTW, and it is computed as the proportion of units in each stratum that are in the treated group in FM, thereby reweighting the HC0 and HC1 subjects to be representative of all the subjects in the current (MORPHEUS-UC) trial. The weighting scheme is similar in estimating the ATE on the treated group, where we regard the current trial population as the treated group. [21,24]

We further explore both IPTW and FM with two modifications (Table 1): separate, SIPTW and SFM; and joint, JIPTW and JFM. The main distinction of these two modifications is in how the weights are obtained. Separate propensity score methods take into account the heterogeneity of the historical controls from the two HTD in two stages with weights estimated based on the study populations: MT and HC0, and MT and HC1, separately. The joint propensity score methods take the heterogeneity of MT, HC0 and HC1 together into account in the final stage of estimation using weighted regression: They include a three-level categorical variable, Q, indicating the source of the data (i.e., MT, HC0 or HC1).

In separate propensity score methods, the weights of HC0 and HC1 are derived separately by two propensity score models, which result in balanced covariate distributions within 1) combined MT and HC0, and 2) combined MT and HC1. However, the covariate distribution balance may not be optimal in the overall combined population after weighting or matching. In contrast, this issue may be addressed when we perform JIPTW or JFM among the joint population of MT and HC0 and HC1, because the joint propensity score methods estimate their propensity scores within the overall population so that the overall covariate distribution balance is better achieved. Note that in the setting where only one historical trial is being incorporated, the separate and joint methods degrade to FM and IPTW.

Previous work by Stuart and Rubin[36] proposed a theoretical framework that includes a matching method to adjust for differences when there are multiple external data sources, with large similar sample sizes of current and historical control arms. In contrast, the current trial here is representative of the smaller sample sizes of a typical Phase Ib/II clinical trial. In the current paper we also evaluated a plausible situation of different levels of heterogeneity between the two HTD. In addition, implementing their method includes recursive procedures and choosing hyperparameters (such as an optimal number of matched samples in one of the control arms), which no software or standard package can yet perform. Yuan et al.[37] proposed a variation to the former work that accommodates smaller sample sizes but requires the experimental treatment arm to have a larger sample size than the control, which again was not the case in part of our scenario. Thus, the current paper aims to contribute to current knowledge about frequentist approaches that can combine control patients' data from two external data sources.

The frequentist methods IPTW and matching are the most popular methods implemented in the literature and draw most of the attention of the current research. Other more complicated frequentist approaches such as doubly robust methods[7-10] are beyond the scope of this work because, in this paper, we aim to understand the performance of two propensity score methods in our setting first.

## 2.2 Commensurate priors within Bayesian dynamic borrowing

Bayesian approaches provide another perspective on how to integrate the HTD. For instance, dynamic borrowing and the commensurate prior approach[38] can employ Bayesian hierarchical modeling where an explicit parameter measuring the variation between studies (here MORPHEUS-UC, HC0 and HC1) is assumed and follows a prior distribution. More precisely, the commensurate prior indicates the prior distribution of the precision (inverse of the variance) or, in other words, the similarity between the current and historical controls. Different borrowing behavior can be achieved by varying the hyperprior. In this paper, we mainly assess the impact of modifications of three commensurate priors within dynamic borrowing on their estimation performance. Several other Bayesian borrowing methods have been proposed for multiple historical data sources, including modified power prior,[39,40] meta-analytic–predictive prior (MAP),[41] Robust MAP,[42] and Bayesian semiparametric MAP.[43] However, caution should be applied in using these approaches when the number of external data sources is limited. In this work, we focus on the commensurate prior approach in Bayesian dynamic borrowing;other potential approaches could be explored in future research. Denote the subject index by $i$, the outcome by $Y_i$, the mean outcomes by $f(\mu_i)$, where $f(\cdot)$ depends on the type of outcome which we specify more analytically in the simulation section, and the estimand by $\delta$.

Depending on whether we regard the two historical trials as different sources of data or not, we can specify the mean outcomes in the following two ways:

$$1. f(\mu_i) = \alpha_0 + \delta, i \in MT, f(\mu_i) = \alpha_0, i \in MC, f(\mu_i) = \alpha_1, i \in HC0, f(\mu_i) = \alpha_2, i \in HC1. \quad (1)$$

or

$$2. f(\mu_i) = \alpha_0 + \delta, i \in MT, f(\mu_i) = \alpha_0, i \in MC, f(\mu_i) = \alpha_3, i \in HC0/HC1. \quad (2)$$

By *HC0/HC1* we mean that the control subject comes from either historical trial without distinguishing which one.

We assume vague priors for $\alpha_0$ and $\delta$ : $\alpha_j \sim N(0, 1000), j = 1,2,3$, $\delta \sim N(0,1000)$. The commensurate approach for the current control arm effect assumes

$\alpha_0 | \alpha_1, \alpha_2 \sim N(\frac{nHC0}{nHC0+nHC1}\alpha_1 + \frac{nHC1}{nHC0+nHC1}\alpha_2, 1/\tau)$, aligned with (1), and

$\alpha_0 | \alpha_3 \sim N(\alpha_3, 1/\tau)$, aligned with (2).

The outcome contrast between experimental treatment and control arms among the population of the current trial is estimated by $\delta$.

The three priors we consider for the parameter $\tau$ are: 1) the Gamma distribution with shape parameter 1 and rate parameter 0.001, referred to as an informative prior. It tends to borrow more historical controls into the estimation, similar to the frequentist methods specifying higher weights for those subjects in the analysis; 2) the Gamma distribution with shape parameter 0.001 and rate parameter 0.001, referred to as a noninformative prior. It tends to borrow less historical controls into the estimation, similar to specifying lower weights in the frequentist methods for those subjects; and 3) the Half-Cauchy distribution of $\sqrt{\tau}$ with center 0 and scale 2.5 (truncated by 0), being referred to as weakly informative prior as default.[44,45] It might be more flexible compared to the two aforementioned priors, as it tends to let the data decide the strength of borrowing.[45] We respectively refer to these three priors as Noninformative Priors Same (NPS), Informative Priors Same (IPS), and Weakly informative Priors Same (WPS). Such methodologies assume no heterogeneity between the two HTD sources. We subsequently account for the heterogeneity between the two HTD sources with distinct priors, following the same distribution, for each of HC0 and HC1, thereby mimicking the frequentist methods with how the weights are obtained when integrating two HTD sources (i.e., through the separate and joint modifications for both IPTW and matching). More precisely, Noninformative Priors Distinct (NPD), Informative Priors

Distinct (IPD), and Weakly informative Priors Distinct (WPD) adjust the heterogeneity by specifying the same prior distribution to the commensurate parameter of HC0 and MC and to the commensurate parameter of HC1 and MC. Therefore, when only one external data source is being incorporated, the approaches with one prior degrade to the approaches with the distinct priors for the two external trials.

Finally, the six examined modifications in the priors as well as two benchmarks (nonborrowing and full-borrowing) are outlined in Table 2. Unlike the considered frequentist approaches, which determine the contribution of each external control with weights derived from the propensity score, the Bayesian commensurate prior defines the borrowing strength through the specification of priors and directly compares the outcomes among different trials. In the examined Bayesian models being still interested in the ATE (a marginal effect) among the subjects in the current trial, we do not incorporate the covariate adjustment. We elaborate more in the Simulation section.

## 3. SIMULATION

### 3.1 Sample size

The sample size of each arm is shown in Figure 1.

As mentioned in the Introduction section, we generated data sets that 1) have the same sample size as MORPHEUS-UC in the preliminary stage, and 2) have 40 subjects (rather than 16) in MT (a plausible number of treated patients in MORPHEUS UC in the secondary stage) and all others the same to imitate the numbers of eligible UC control patients from HC0 and HC1.

### 3.2 Covariates

We considered thirteen baseline (both continuous and binary) covariates denoted by **X** (since this was the number of the key baseline covariates that were available and common both in the MORPHEUS-UC and the two HTD trials). These covariates were generated according to summary statistics of the real data in each arm, and further details can be found in the R code. We assumed (in the propensity score methods) that covariates are balanced in MT and MC, but not balanced (i.e., they had different means) between current and historical controls. We did not examine or take into account correlation between these thirteen variables. The same data generating mechanism for covariates was used across all the simulations.

One of these covariates ($X_8$) was considered as a significant treatment effect modifier (or predictive biomarker, based on clinical knowledge for how these drugs are expected to work).

### 3.3 Treatment arm assignment

We further created two nested categorical variables D and A to represent the subjects' treatment arm among the three data sources. First, arm variable D is a four-level categorical variable (MT/MC/HC0/HC1). Second, treatment variable A is 1 if the subject was in MT, representing that the subject was assigned to the experimental treatment arm in the current trial (atezo combo), it is 0 otherwise (atezo mono). Both D and A were generated independently of the 13 covariates. The reference level of D is MC. Note that the covariates distributions are different among different data sources.

### 3.4 Conditional trial-specific effects

We next specify the conditional trial-specific arm effects. We aim to evaluate estimators' performances in different scenarios. We assumed the conditional effect of the experimental treatment (atezo combo) compared to controls (atezo mono) can be similar or different among the three data sources. The conditional effects contrast is denoted by $\boldsymbol{\beta} = (\beta_1, \beta_2, \beta_3)$, where $\beta_1, \beta_2, \beta_3$ are the conditional log HRs between a) MT and MC, b) HC0 and MC, and c) HC1 and MC, respectively. Without loss of generality, we evaluated four simulation scenarios (Table 3).

The difference of $\beta_1, \beta_2, \beta_3$ may be because of the different effects between the treatment (atezo combo) and controls (atezo mono) as well as the heterogeneity of population characteristics from the different trials. Scenario 1 represents the case where the conditional treatment effect is the same across all arms among the three data sources. Scenario 2 assumes a larger treatment effect in the treatment arm compared to all controls (historical and current; in other words, there are no differences among the controls). Scenario 3 is the scenario whereas in Scenario 2, a larger treatment effect in the treatment arm compared to all controls is assumed, but the treatment effects are different among comparisons of the hybrid controls. More precisely, in Scenario 3, it is assumed that there are no differences between historical controls from the two HTD, but there are differences between current and historical controls (for example, in Scenario 3, $\boldsymbol{\beta} = (\log(0.5), \log(3), \log(3))$. Scenario 4 is the case where the treatment effects compared to hybrid controls are different in the three trials. In the simulation, the survival time was generated based on the log HR of the treatment effect conditional on the baseline covariates.

### 3.5 Outcome

We use survival outcomes in this work. We assumed that the survival event times follow a Weibull distribution with scale parameter $exp(\eta_i)$ and shape parameter $p$ where $i$ denotes a subject, $\eta_i$ is $\eta_i = \beta_0 + \boldsymbol{\beta} \cdot D_i + \gamma \cdot X_i + \xi \cdot X_{8i} \cdot A_i$. Denote the survival event time by $T$, $T_i \sim WB(exp(\eta_i), p)$, generated by inverting the survival distribution, represented by the following equation, $T_i = -[\frac{\log(U)}{exp(\eta_i)}]^{1/p}$, where in the simulation, U is uniformly distributed ranging from 0 to 1, $p$ is set to 2. Generating the event time in such a way indicates the corresponding hazard function $h(t_i) = pt^{p-1}exp(\eta_i)$, and the log event time follows a Gumbel distribution with location $\eta_i$ and scale $1/p$, which satisfies the assumptions of both models used here, Cox proportional hazards model and accelerated failure time (AFT).[46]

The covariates' coefficients $\gamma$ were set from 0.01 to 0.07, increased by 0.005 to represent the different magnitude of covariate effects on the outcome. The coefficient of the interaction of $X_8$ and treatment, $\xi$ was set to 0.01.

Censoring times in each arm were generated to follow normal distributions,[2] where the mean is 1.5 times the mean of event times in each arm, and the variance of 0.1. If the censoring time was less than the event time, then the subject was censored. Thus, the survival times were the minimum of censoring times and event times. The choice of normal distribution of the censoring time was supported by the real data. The distribution of the survival times of the real data follows a similar distribution with the distribution of the minimum simulated survival time and censoring time.

## 3.6 Causal Estimand

The target parameter was the ATE, which is the contrast effect of the experimental treatment (atezo combo) and control (atezo mono) arms in the population represented by current (MORPHEUS-UC) trial. With a survival outcome, a natural target parameter is the marginal log HR of MT versus MC.

The conditional log HR is not equivalent to the marginal log HR due to noncollapsibility.[47] Therefore, for each simulation scenario, we generated the marginal log HR (true values of targeted parameters). First, we sample (with replacement) 200 indices from MORPHEUS-UC data, both MT and MC subjects. Next, we generate covariates and two counterfactual outcomes (one by assuming all subjects received the experimental treatment, and the second assuming all subjects received the control) for subjects whose indices were generated in step one. Subsequently, we combine the data resulting in 400 samples, with each indexed subject accounting for at least two samples. Then we regress the counterfactual outcome on the assumed treatment assignment. Finally, we obtain the coefficient of the treatment parameter. We repeated this procedure 1,000,000 times and took the mean of the coefficients in each repetition as the true value of the target parameter.

## 3.7 Evaluation metrics

We evaluated an estimator's performance by its bias, variance, mean squared error (MSE) and effective sample size (ESS). Explicit power and type I error considerations for a hypothesis test are not included in the MORPHEUS-UC design as is typical of Phase Ib/II signal-seeking studies; therefore, they were not assessed in our simulation studies.

Denote the number of repetitions for each simulation study, the true value, and the estimate by $N_{sim}$, $\theta_0$, and $\hat{\theta}$, respectively, and $\bar{\theta} = \frac{1}{N_{Sim}} \sum_{j=1}^{N_{Sim}} \hat{\theta}_j$, as the mean of the estimates from all repetitions. Then bias, variance, and MSE are computed as $|\bar{\theta} - \theta_0|$, $\frac{1}{N_{sim}-1} \sum_{j=1}^{N_{Sim}} (\hat{\theta}_j - \bar{\theta})^2$ and $\frac{1}{N_{sim}} \sum_{j=1}^{N_{sim}} (\hat{\theta}_j - \theta_0)^2$, respectively.

To calculate the ESS for the frequentist methods, we use the methods provided by McCaffrey et al.[48, 49]. Essentially, the ESS is obtained by the square of the sum of the weights of HTD divided by the sum of the square of the weights. The method is similar to the one used in the R packages such as MatchIt[33] and optweight[50]. The ESS given by this method provides an estimate of the number of individuals in HTD comparable with the current trial group. For the Bayesian methods, we calculate the ESS using the variance and precision ratio method[51]. The ESS can be obtained by the total sample size times the ratio of the variance of the estimator and prior. Other methods to calculate the ESS exist[51]; in this work, we use the above methods for illustration.

In total, we had four scenarios of conditional effects multiplied by two sample sizes of MT, yielding eight settings to be evaluated for each of the considered methods.

### 3.8 Method implementation

As mentioned before, the bias, variance, and MSE were used to evaluate the performance of the methods introduced in Section 2. In all the examined frequentist approaches, propensity scores were modeled by logistic regression. Within this logistic regression, we excluded the interaction term ($X_{8i} \cdot A_i$) to represent model misspecification. We also excluded from this logistic regression three covariates with coefficients 0.06, 0.065, and 0.07 to represent the issue of unmeasured confounding.

Furthermore, with the frequentist approaches, we used AFT (Frequentist Weibull) and Cox proportional hazard models (Frequentist Cox) to estimate the marginal log HR. As the coefficients in the AFT models cannot be directly interpreted as the log HR, we convert those to the log HR by dividing them by the corresponding estimated scale parameter. With Bayesian dynamic borrowing, we used a Weibull distribution to model survival times and the zero-trick[52] to specify the model likelihood.

Investigating these three survival models is a fair comparison since the simulations generated outcomes by Weibull distribution, which satisfies the assumptions of both AFT and Cox models. We repeated each of the simulations 500 times ($N_{sim}$). The bias, variance, and MSE are calculated from the resulting point estimates within each simulation scenario for each of the methods. In Bayesian analysis, the number of MCMC chains, iterations in the adaptive phase, and iterations in the sampling phase were set to 3, 1000, and 2000, respectively. We use the mean to summarize the posterior distribution of the parameters.

The code (both in R and jags) is provided at https://github.com/phcanalytics/HybridControl_early.

## 4. RESULTS

The biases, variances, and MSEs of the estimators from the various examined methods with varying sample sizes of MT, n=16 and n=40, are shown in Figures 2 and 3, respectively. The actual numerical results can also be found in the Appendix (Appendix Tables 1 and Table 2). Simulation scenarios are also described in Table 3.

Focusing on the bias, in the frequentist approaches, all four modifications showed similar performance in Scenarios 1 and 2, but joint propensity score methods outperformed separate ones in Scenarios 3 and 4. The phenomenon was more significant when the AFT model was used, and the difference was larger when MT=40 in comparison to MT=16 (Figures 1B and 2B). In addition, JIPTW generated less bias than JFM, but the difference is negligible. When the AFT model was used, SFM had more bias than SIPTW. In commensurate prior within Bayesian dynamic borrowing with no adjustment for covariates, in Scenarios 3 and 4, noninformative priors were associated with the least bias compared to all other approaches regardless of sample size, while for Scenario 1 the differences were minimal and for Scenario 2 the differences were less distinct. Weakly informative priors generated similar biases to noninformative priors, but informative priors were largely biased. The difference between NPD and NPS is negligible, and the same for IPD and IPS, as well as for WPD and WPS. In general, the least biased Bayesian approaches were less biased than the frequentist ones, except in Scenario 4 with a larger sample size, where JIPTW AFT was better than the Bayesian NPS.

In terms of variance, the differences among each of the explored methods were minimal across the four simulation scenarios regardless of the sample size of the experimental treatment arm, but the larger MT=40 sample size reduced the variances in general. Although the differences were minimal, in the frequentist approaches, the variances produced from the joint propensity score methods exceeded those from the separate propensity score methods and in commensurate prior within Bayesian dynamic borrowing, full-borrowing with no adjustment on covariates generating the least.

Regarding MSE, as a trade-off between bias and variance: All the examined methods, regardless of the sample size of MT, show similar performance in the more homogeneous treatment effects (i.e., Scenarios 1 and 2). In Scenarios 3 and 4, separate propensity score methods, as well as Bayesian full-borrowing, nonborrowing, and informative priors with no adjustment on covariates, produced larger MSE compared to other methods where MSE is similar. In general, the joint propensity score methods slightly outperformed the examined Bayesian dynamic borrowing (by using noninformative priors) with no adjustment on covariates.

Concerning ESS, the commensurate prior within Bayesian dynamic borrowing, using the weakly-informative commensurate priors, reached the largest ESS, followed by the non-informative priors, and then the informative ones. In the frequentist methods, the joint propensity score methods generated larger ESS than the separated propensity score approaches. In general, the frequentist methods outperformed the commensurate prior within Bayesian dynamic borrowing with no adjustment on covariates, with larger ESS.

## 5. DISCUSSION

In this paper, we explored hybrid control methods from both the frequentist and Bayesian frameworks, tailored for estimating ATE in early-phase oncology trials with a time-to-event outcome and small sample sizes. In particular, the investigated hybrid control methods can be used when two HTD sources are available and relevant HTD controls' data are eligible to be integrated. Furthermore, we employed comprehensive simulations to assess the robustness of such methods in situations reflecting clinical trial practice with various scenarios such as the presence of model misspecification, unmeasured confounding, different sample sizes of current treatment groups, and magnitudes of treatment effect, as well as varying heterogeneity among the three trials via extensive comparisons of estimation performances in terms of bias, variance, and MSE.

Within the frequentist approaches, in addition to SFM or SIPTW, we also proposed and explored JFM and JIPTW, which is, to our knowledge, a novelty of our work. The joint propensity score methods allowed for the treatment effects to vary between the different trials, as we included the trial variable Q in the relevant weighted regressions. With Bayesian dynamic borrowing, we explored variations of commensurate priors as well as the impact of specifying two parameters with the same prior distribution (distinct) and one prior (same) for the assumption of lack of heterogeneity between the historical controls from the two HTD sources. Associated results did not provide evidence of much difference in the distinct versus same priors for the comparison of current controls and historical controls from the two HTD. Nevertheless, the careful consideration of different prior distributions of the commensurate parameters in the Bayesian dynamic borrowing is important in analyzing the data as shown from our simulation results. More detailed discussion about prior distributions for variance parameters in Bayesian hierarchical models can be found in Gelman (2006).[44] When there are available historical controls from only one HTD, the methods mentioned previously degrade to simplified methods such as specifying only one commensurate prior in Bayesian dynamic borrowing or not regressing on the additional variable trial Q in the frequentist approaches. When there are more than two HTD, methods such as MAP[41] and its different forms[42,43] can be used.

With reference to bias, for the less heterogeneous treatment effect scenarios, we did not find any major unexpected differences among the explored methods. For instance, for Scenario 2, where perfectly commensurate historical data are assumed, it is intuitive to have large bias when no borrowing from historical data is assumed. For the more heterogeneous treatment effect scenarios, overall, Bayesian dynamic borrowing with noninformative priors for the historical controls from the two HTD was less biased than joint propensity score methods. Within the examined frequentist approaches, JIPTW was slightly less biased than JFM, and we also observed that the weights produced from IPTW were usually smaller than FM in both separate and joint propensity score methods. Specifically, fewer historical controls from the two HTD would be integrated with IPTW, which is more desirable when there is more heterogeneity between current and historical trials. However, previous research[31,32] suggests that matching is more robust to the propensity score model misspecification than weighting in large sample size settings. Explanations for the IPTW outperforming FM in our study are possibly the small sample size of the treatment arms with different degrees of heterogeneity among the three trials and the way we have defined model misspecification (i.e., omitting the effect modification as well as considering unmeasured confounding by omitting three variables that generated the outcome). In a previous research model misspecification was defined as the violation of parametric assumption of the propensity score model (i.e., there is a linear relationship between the logit of treatment assignment and covariates). Furthermore, the AFT model is a parametric model with two additional parameters (shape and scale) to be estimated and thus is more sensitive to the model specification compared to the Cox model as a semiparametric model. This is the reason separate propensity score methods were largely biased when the AFT model was used when the heterogeneity is large (e.g., in Scenario 3 and 4). Further research can be done with the use of the trial variable Q to avoid model misspecification. In conclusion, based on the data we generated, to obtain the least bias, we suggest JIPTW and JFM from the explored frequentist approaches as well as NPS or NPD within the commensurate prior within Bayesian dynamic borrowing models without covariates.

Concerning variance, within the investigated commensurate prior within Bayesian dynamic borrowing with no adjustment on covariates, as expected, depending on the increased strength of borrowing, the variance decreased. For instance, full-borrowing makes full use of the data yielded a smaller variance. In general, we saw reverse performance between bias and variance produced from each method for the most heterogeneous treatment effect scenarios. This makes sense, as in such situations fewer data from historical trials were borrowed, which reduces the precision of the estimates.

Consequently, larger MSEs in all explored methods were driven mainly by their more pronounced biased estimates. In general, the MSEs from the examined frequentist approaches and the commensurate prior within Bayesian dynamic borrowing with noninformative and weakly informative priors were similar. The joint propensity score methods had slightly smaller MSEs than the Bayesian in the most heterogeneous scenarios, possibly due to not including the covariates into the Bayesian models.

Regarding the ESS of commensurate prior within Bayesian dynamic borrowing, since we assumed the same prior distribution of the parameter of interest ($\delta$) regardless of the commensurate priors used, the ESS essentially depends on the variance of the posterior distribution of $\delta$. Thus, weakly-informative priors, which are half-Cauchy distributed, allowed more space for data to decide the level of borrowing, resulting in a more diffused posterior distribution, and therefore attained a larger ESS. In the frequentist methods, joint propensity score methods created larger ESS. This is because they estimated the weights using the two HTD together, resulting in a less variable set of weights (compared to the separated methods). The ESS of frequentist methods were larger than Bayesian ones, possibly because the Bayesian methods ignored the covariate adjustment. Generally, the smaller the biases, the larger the ESS. The variances of the estimators also impact the ESS, for instance, the larger ESS of Bayesian methods, the larger variances of the estimators. Similar to MSE, the ESS can also be regarded as a bias-variance tradeoff, though from our observations, the ESS is dominated by the biases.

RExtensions of current research and future work might include exploring a wider range of priors. For instance, one could specify another noninformative prior that follows a Gamma distribution with shape parameter 0.001 and rate parameter $10^{-5}$ and examine its impact on estimation performance. Another interesting aspect would be to explore whether 2 stage Bayesian approaches (i.e., first pair matching on patients with propensity scores and then using Bayesian dynamic borrowing to select among those with different commensurate priors) provide better performance. This is beyond the scope of this paper, but in such an approach, the heterogeneity would be modeled in the first stage via baseline covariates in the propensity score matching as well as via the selected priors in the second stage, and thus the acquired causal estimand becomes less clear. In other words, in such an approach the causal estimand would correspond to the ATE in the treated arm sample after propensity score matching and the sample of historical controls via the selected priors that the Bayesian model eventually retained; such an estimand might not correspond to any specific target population.

Another aspect for future research with commensurate prior within Bayesian dynamic borrowing to consider is the noncollapsibility issue when the acquired estimand is the marginal effect; the explored method must be able to estimate marginal rather than conditional HRs.[53-55]

Moreover, in this work we focus on the ATE estimation and evaluation. More efforts can be made to develop and evaluate various methods when the estimand is the heterogeneous treatment effect (HR conditional on the potential effect modifiers, $X_8$). Estimating such effects in early phase trials where sample sizes tend to be small can be challenging as a large sample size is required, especially when the covariates to be conditioned on are continuous or have a large number of categories. We leave this research as future work.

In the simulation, it is not surprising that when all causal assumptions are satisfied, and the treatment effects are the same across all the source data (scenario 1), the frequentist methods outperformed Bayesian ones. This is because the propensity score adjustment incorporated in the frequentist methods corresponds to the data generation mechanism. In contrast, Bayesian methods only adjust for confounding in a data-driven way. In practice, where the data generation mechanism is unknown, the results may not be extrapolated.

While this work empirically verified and rationalized the comparison among frequentist and Bayesian methods in different scenarios, we did not extend the theory development. For instance, the variance of the estimators of the frequentist methods can be left for future work.

Lastly, all the simulation results are based on the data generated similar to the real data of the MORPHEUS-UC trial and the two IMvigor trials. We encourage readers to examine the trial operating characteristics based on their own trial setting, and our work can serve as a reference framework. The R code of our simulation is also publicly available and can easily be modified or incorporated in the building of web applications (i.e., Shiny apps) that enables more users to input the covariate distributions according to their applications in early-phase oncology trials to further compare the obtained results as well as further assess the potential for their generalization. Such practices can guide the confirmation of selecting appropriate methods to analyze hybrid control trials in similar settings.

Overall, our simulations with different magnitudes of treatment effect in the current trial, different levels of heterogeneity among the three trials, and varying sample sizes of current treatment groups indicated that various methods perform differently in different settings. There were fewer differences when the treatment effects across the three trials are considered smaller and more homogeneous. With more assumed differences in treatment effects, joint propensity score methods and weakly informative as well as noninformative priors used with Bayesian dynamic borrowing with no adjustment on covariates are recommended. For generalization of hybrid control methods in such settings we recommend more simulation studies where different magnitudes of treatment effects can also be tested.

## APPENDIX

The appendix contains Appendix Table 1 and 2, which are the numeric results of the simulations.

**Funding Information:** The work is fully funded by Roche

**Author contributions:** All authors had full access to all the data in the study and take responsibility for the integrity of the data and the accuracy of the data analysis. All authors contributed substantially to the study design, data analysis and interpretation, and writing of the manuscript.

## CONFLICTS OF INTEREST

Guanbo Wang: employee of in F. Hoffmann-La Roche Ltd

Melanie Poulin Costello: employee of and owns stock in F. Hoffmann-La Roche Ltd

Herbert Pang: employee of Genentech, Inc. and owns stock in F. Hoffmann-La Roche Ltd

Jiawen Zhu: employee of Genentech, Inc. and owns stock in F. Hoffmann-La Roche Ltd

Hans-Joachim Helms: employee of and owns stock in F. Hoffmann-La Roche Ltd

Irmarie Reyes-Rivera: employee of and owns stock in F. Hoffmann-La Roche Ltd

Robert W. Platt: no conflict of interest

Menglan Pang: employee of Biogen Inc.

Artemis Koukounari: employee of F. Hoffmann-La Roche Ltd

**Data sharing statement:** Qualified researchers may request access to individual patient-level data through the clinical study data request platform (https://vivli.org/). Further details on Roche's criteria for eligible studies are available here (https://vivli.org/members/ourmembers/). For further details on Roche's Global Policy on the Sharing of Clinical Information and how to request access to related clinical study documents, see here (https://www.roche.com/research_and_development/who_we_are_how_we_work/clinical_trials/our_commitment_to_data_sharing.htm). This is a methods paper includes R code which is available on open source.

## SUPPORTING INFORMATION

Additional supporting information may be found in the online version of the article at the publisher's website. The code (both in R and jags) is provided at https://github.com/phcanalytics/HybridControl_early.

Table 1: Description of the two examined frequentist approaches (matching and weighting) and their two modifications 1) separate and 2) joint

| Methods name | Description |
| --- | --- |
| 1. SFM (Separate Full Matching*) | MT and HC0 are matched by FM; MT and HC1 are matched by FM. Weights of HC0 and HC1 are obtained from those matchings separately. Weighted regress outcomes on the treatment. |
| 2. SIPTW (Separate IPTW) | Propensity scores of HC0 are obtained by modeling the probability of the subject receiving the treatment conditional on covariates among the subject of MT and HC0. Similar to the propensity score of HC1. Weighted regress outcomes on the treatment. |
| 3. JFM (Joint Full Matching*) | Perform an FM among MT, HC0, and HC1 to obtain the weights of HC0 and HC1. In the weighted regression, include the variable trial Q as a covariate apart from the treatment. |
| 4. JIPTW (Joint IPTW) | Perform an IPTW among MT, HC0, and HC1 to obtain the weights of HC0 and HC1. In the weighted regression, include the variable trial Q as a covariate apart from the treatment. |

* also referred to as optimal full matching

Table 2: Description of the three examined priors in the Bayesian dynamic borrowing and their two modifications 1) Distinct and 2) Same, as well as two benchmarks

| Priors for the precision parameter in the Bayesian Borrowing approach | Description |
|---|---|
| 1. NPD (Noninformative Priors Distinct) | Specify **distinct** noninformative priors for $\tau$ |
| 2. NPS (Noninformative Prior Same) | Specify the **same** noninformative prior for $\tau$ |
| 3. IPD (Informative Priors Distinct) | Specify **distinct** informative priors for $\tau$ |
| 4. IPS (Informative Prior Same) | Specify the **same** informative prior for $\tau$ |
| 5. WPD (Weakly informative Priors Distinct) | Specify **distinct** weakly informative priors for $\sqrt{\tau}$ |
| 6. WPS (Weakly informative Priors Same) | Specify the **same** weakly informative prior for $\sqrt{\tau}$ |
| 7. NB (Nonborrowing) | **Benchmark** of no information from the historical control but only the MORPHEUS-UC trial. |
| 8. FB (Full-Borrowing) | **Benchmark** to fully merge historical controls into current controls. |

Table 3: Simulation scenarios: specifications of conditional hazard ratios of different control arms versus treatment

| Scenario (effect comparison) | Conditional hazard ratios |
|---|---|
| 1 (MT=MC=HC1=HC0) | MT/MC=HC1/MC=HC0/MC=1 |
| 2 (MT>MC=HC1=HC0) | MT/MC=0.5, HC1/MC=HC0/MC=1 |
| 3 (MT>MC>HC1=HC0) | MT/MC=0.5, HC1/MC=HC0/MC=3 |
| 4 (MT>MC>HC1>HC0) | MT/MC=0.5, HC1/MC=3, HC0/MC=12 |

Figure 1: Sample sizes of each arm in the MORPHEUS-UC trial and historical controls. Displayed numbers represent sample sizes in the MORPHEUS-UC trial in the preliminary and secondary stages as well as numbers of eligible UC control patients from the two IMvigor trials.

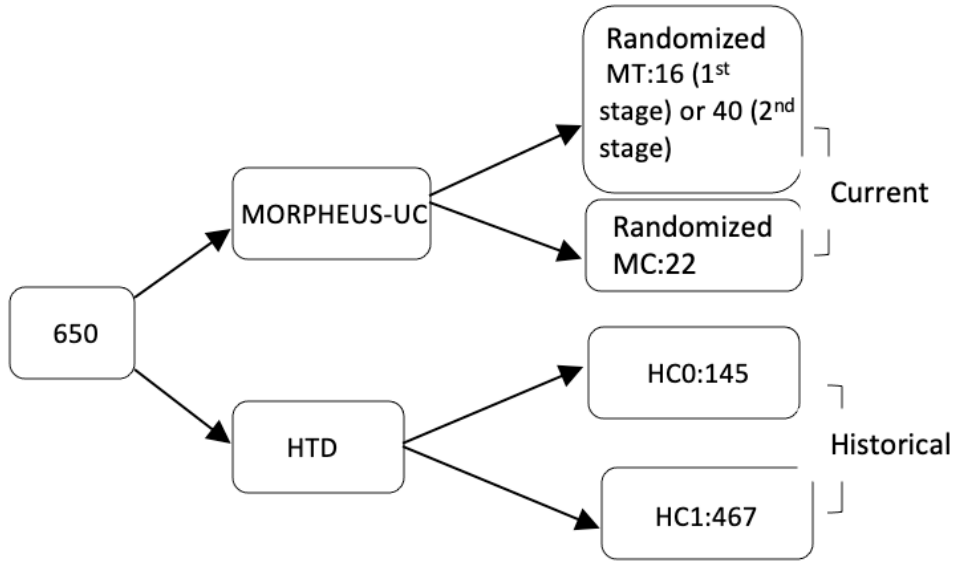

Figure 2: Simulation results when the sample size of MT is 16. (A)-(C) are the resulted biases from different methods, (D)-(F) are the resulted variance from different methods, (G)-(I) are the resulted MSE from different methods, and (J)-(L) are the resulted ESS from different methods.

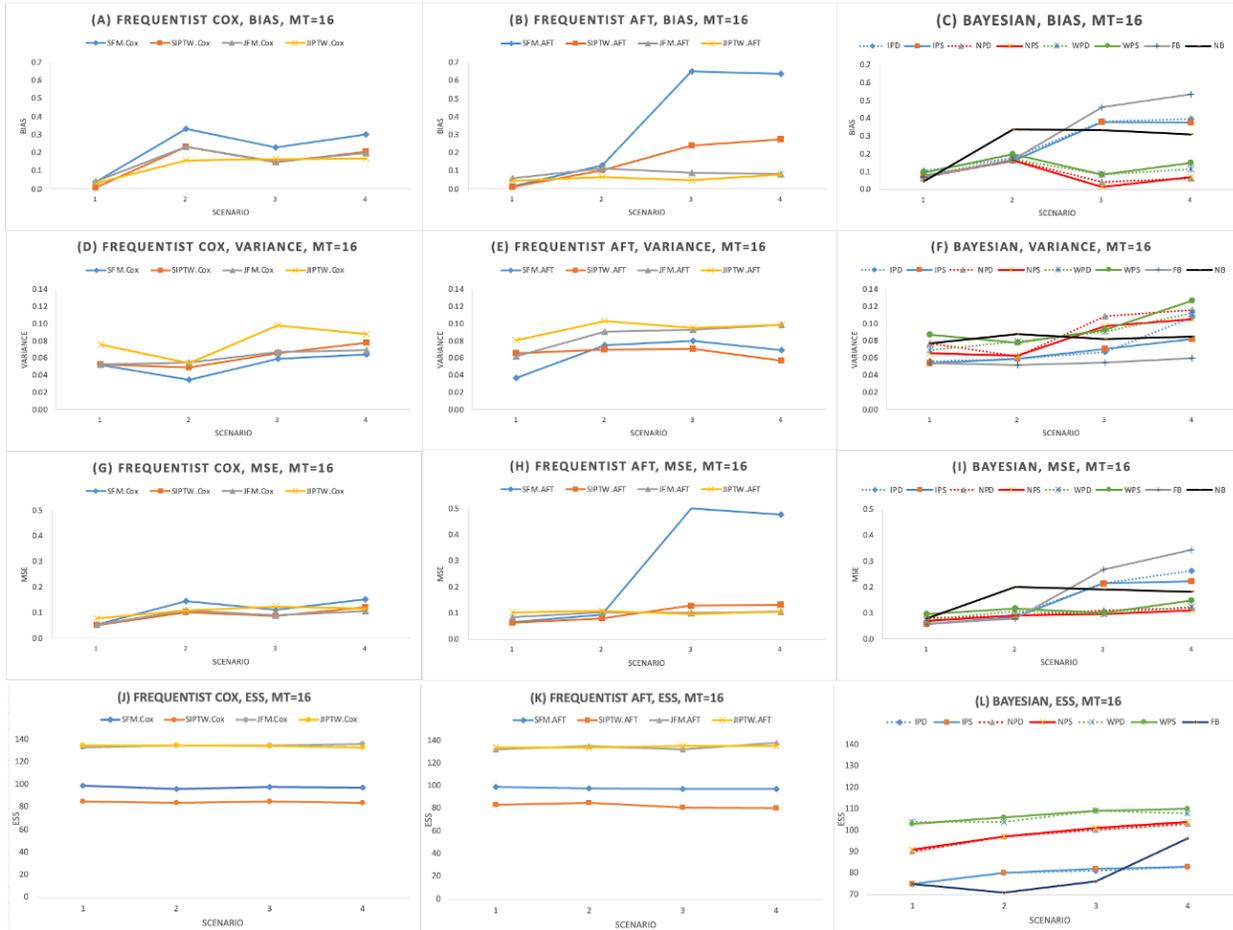

Figure 3: Simulation results when the sample size of MT is 40. (A)-(C) are the resulted biases from different methods, (D)-(F) are the resulted variance from different methods, and (G)-(I) are the resulted MSE from different methods, and (J)-(L) are the resulted ESS from different methods.

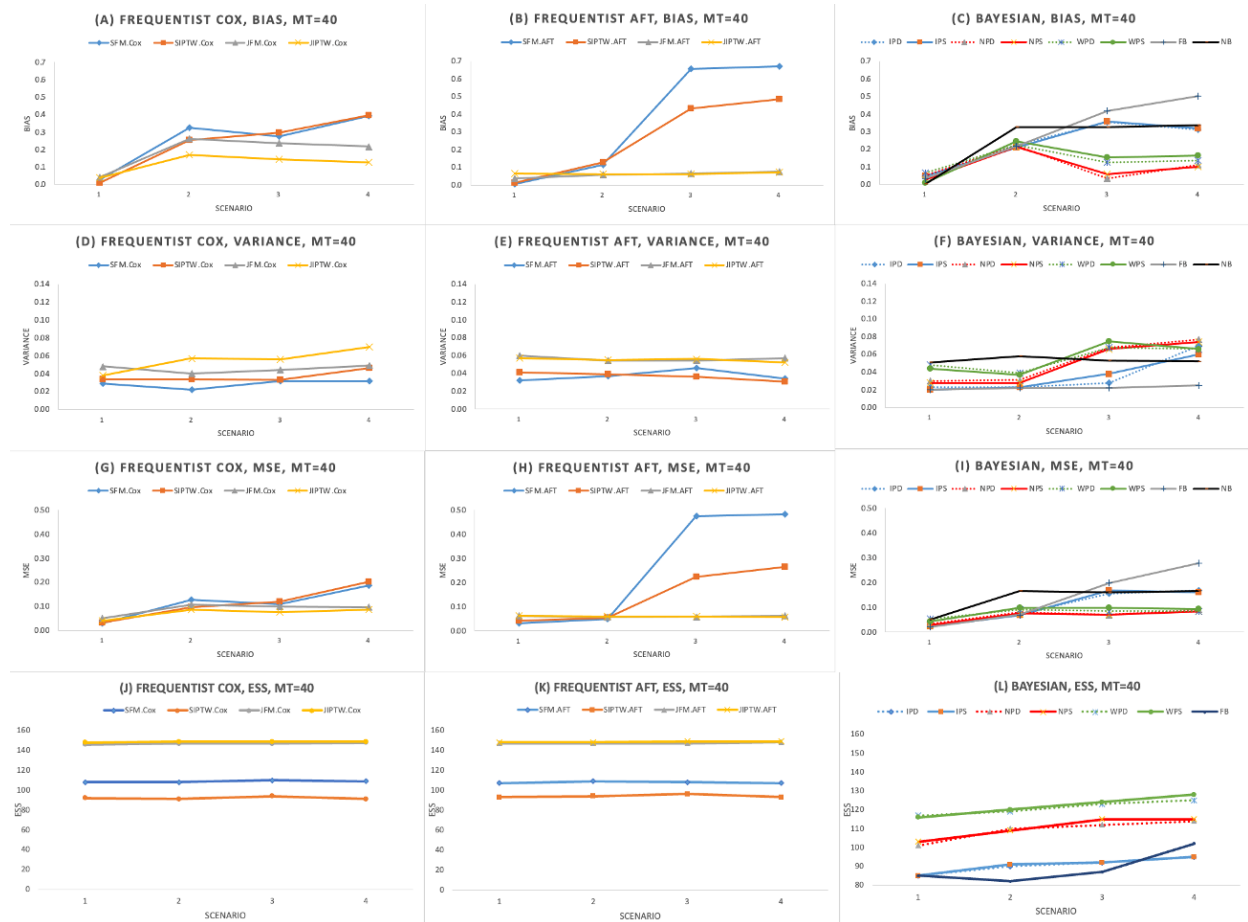

**Appendix:** Numerical results

Appendix Table 1: Simulation results when the sample size of MT is 16. The gray cells refer to the methods that have the least bias/variance/MSE/ESS.

Appendix Table 2: Simulation results when the sample size of MT is 40. The gray cells refer to the methods that have the least bias/variance/MSE/ESS.

### MT=16

| Scenario | 1.MT=MC=HC1=HC0 | | | | 2.MT>MC=HC1=HC0 | | | | 2.MT>MC>HC1=HC0 | | | | 4.MT>MC>HC1>HC0 | | | |
|---|---|---|---|---|---|---|---|---|---|---|---|---|---|---|---|---|
| Metric | Bias | Variance | MSE | ESS | Bias | Variance | MSE | ESS | Bias | Variance | MSE | ESS | Bias | Variance | MSE | ESS |
| Method | Bayesian Weibull | | | | | | | | | | | | | | | |
| IPD | 0.104 | 0.056 | 0.067 | 75 | 0.176 | 0.059 | 0.090 | 80 | 0.384 | 0.067 | 0.215 | 81 | 0.396 | 0.107 | 0.264 | 83 |
| IPS | 0.075 | 0.054 | 0.059 | 75 | 0.159 | 0.059 | 0.084 | 80 | 0.379 | 0.071 | 0.214 | 82 | 0.375 | 0.082 | 0.222 | 83 |
| NPD | 0.066 | 0.077 | 0.081 | 90 | 0.166 | 0.062 | 0.089 | 97 | 0.041 | 0.109 | 0.110 | 100 | 0.064 | 0.116 | 0.119 | 103 |
| NPS | 0.071 | 0.066 | 0.071 | 91 | 0.165 | 0.063 | 0.090 | 97 | 0.013 | 0.097 | 0.097 | 101 | 0.068 | 0.105 | 0.110 | 104 |
| WPD | 0.080 | 0.069 | 0.075 | 104 | 0.169 | 0.079 | 0.107 | 104 | 0.082 | 0.091 | 0.097 | 109 | 0.116 | 0.112 | 0.125 | 108 |
| WPS | 0.093 | 0.087 | 0.095 | 103 | 0.197 | 0.078 | 0.117 | 106 | 0.084 | 0.093 | 0.100 | 109 | 0.148 | 0.127 | 0.148 | 110 |
| FB | 0.070 | 0.054 | 0.059 | 75 | 0.166 | 0.052 | 0.079 | 71 | 0.462 | 0.055 | 0.268 | 76 | 0.534 | 0.060 | 0.345 | 96 |
| NB | 0.044 | 0.077 | 0.079 | - | 0.336 | 0.088 | 0.201 | - | 0.333 | 0.082 | 0.192 | - | 0.311 | 0.085 | 0.182 | - |
| Best for Bayesian | NB | IPS | IPS | WPD | IPS | FB | FB | WPS | NPS | FB | NPS | WPD | NPD | FB | NPS | WPS |
| | Frequentist Weibull (AFT) | | | | | | | | | | | | | | | |
| SFM.AFT | 0.014 | 0.037 | 0.066 | 99 | 0.132 | 0.075 | 0.093 | 98 | 0.650 | 0.080 | 0.502 | 97 | 0.639 | 0.069 | 0.477 | 97 |
| SIPTW.AFT | 0.012 | 0.066 | 0.062 | 83 | 0.105 | 0.070 | 0.080 | 85 | 0.240 | 0.071 | 0.128 | 81 | 0.275 | 0.057 | 0.132 | 80 |
| JFM.AFT | 0.058 | 0.062 | 0.085 | 132 | 0.114 | 0.091 | 0.104 | 135 | 0.091 | 0.093 | 0.101 | 132 | 0.083 | 0.099 | 0.106 | 138 |
| JIPTW.AFT | 0.044 | 0.081 | 0.103 | 134 | 0.064 | 0.103 | 0.107 | 134 | 0.049 | 0.095 | 0.097 | 135 | 0.080 | 0.099 | 0.105 | 135 |
| Best for AFT | SIPTW.AFT | SFM.AFT | SIPTW.AFT | JIPTW.AFT | JIPTW.AFT | SIPTW.AFT | SIPTW.AFT | JFM.AFT | JIPTW.AFT | SIPTW.AFT | JIPTW.AFT | JIPTW.AFT | JIPTW.AFT | SIPTW.AFT | JIPTW.AFT | JFM.AFT |
| | Frequentist (Cox) | | | | | | | | | | | | | | | |
| SFM.Cox | 0.038 | 0.052 | 0.053 | 99 | 0.333 | 0.035 | 0.145 | 96 | 0.229 | 0.059 | 0.111 | 98 | 0.300 | 0.064 | 0.154 | 97 |
| SIPTW.Cox | 0.009 | 0.053 | 0.052 | 85 | 0.234 | 0.049 | 0.103 | 84 | 0.148 | 0.066 | 0.087 | 85 | 0.207 | 0.078 | 0.121 | 84 |
| JFM.Cox | 0.040 | 0.053 | 0.055 | 133 | 0.234 | 0.055 | 0.110 | 135 | 0.150 | 0.067 | 0.090 | 135 | 0.198 | 0.069 | 0.108 | 136 |
| JIPTW.Cox | 0.032 | 0.076 | 0.077 | 135 | 0.159 | 0.054 | 0.109 | 135 | 0.165 | 0.098 | 0.125 | 134 | 0.168 | 0.088 | 0.116 | 133 |
| Best for Cox | SIPTW.Cox | SFM.Cox | SIPTW.Cox | JIPTW.Cox | JIPTW.Cox | SFM.Cox | SIPTW.Cox | JFM.Cox | SIPTW.Cox | SFM.Cox | SIPTW.Cox | JFM.Cox | JIPTW.Cox | SFM.Cox | JFM.Cox | JFM.Cox |
| Best from all methods | SIPTW.Cox | SFM.AFT | SIPTW.Cox | JIPTW.Cox | JIPTW.AFT | SFM.Cox | FB | JFM.AFT | NPS | FB | SIPTW.Cox | JIPTW.AFT | NPD | SIPTW.AFT | JIPTW.AFT | JFM.AFT |

IPD: Informative Prior Distinct; IPS: Informative Prior Same; NPD: Non-informative Prior Distinct; NPS: Non-informative Prior Same; WPD: Weakly-informative Prior Distinct; WPS: Weakly-informative Prior Same;
FB: Full-Borrowing; NB: No Borrowing; SFM: Separate Full Matching; SIPTW: Separate Inverse Probability Treatment Weighting; JFM: Joint Full Matching; JIPTW: Joint Inverse Probability Treatment Weighting

### MT=40

| Scenario | 1.MT=MC=HC1=HC0 | | | | 2.MT>MC=HC1=HC0 | | | | 2.MT>MC>HC1=HC0 | | | | 4.MT>MC>HC1>HC0 | | | |
|---|---|---|---|---|---|---|---|---|---|---|---|---|---|---|---|---|
| Metric | Bias | Variance | MSE | ESS | Bias | Variance | MSE | ESS | Bias | Variance | MSE | ESS | Bias | Variance | MSE | ESS |
| Method | Bayesian Weibull | | | | | | | | | | | | | | | |
| IPD | 0.045 | 0.023 | 0.025 | 85 | 0.223 | 0.023 | 0.072 | 90 | 0.356 | 0.028 | 0.155 | 92 | 0.312 | 0.070 | 0.167 | 95 |
| IPS | 0.049 | 0.020 | 0.023 | 85 | 0.211 | 0.023 | 0.067 | 91 | 0.359 | 0.038 | 0.167 | 92 | 0.319 | 0.060 | 0.161 | 95 |
| NPD | 0.051 | 0.030 | 0.032 | 101 | 0.220 | 0.031 | 0.080 | 110 | 0.036 | 0.068 | 0.069 | 112 | 0.112 | 0.077 | 0.090 | 114 |
| NPS | 0.028 | 0.028 | 0.029 | 103 | 0.214 | 0.028 | 0.074 | 109 | 0.058 | 0.066 | 0.070 | 115 | 0.101 | 0.074 | 0.084 | 115 |
| WPD | 0.066 | 0.048 | 0.052 | 117 | 0.226 | 0.039 | 0.090 | 119 | 0.125 | 0.068 | 0.084 | 123 | 0.136 | 0.066 | 0.084 | 125 |
| WPS | 0.010 | 0.044 | 0.044 | 116 | 0.247 | 0.037 | 0.098 | 120 | 0.153 | 0.075 | 0.098 | 124 | 0.166 | 0.066 | 0.093 | 128 |
| FB | 0.035 | 0.020 | 0.021 | 85 | 0.218 | 0.022 | 0.069 | 82 | 0.419 | 0.022 | 0.198 | 87 | 0.504 | 0.025 | 0.278 | 102 |
| NB | 0.003 | 0.051 | 0.051 | - | 0.328 | 0.058 | 0.165 | - | 0.326 | 0.053 | 0.159 | - | 0.338 | 0.052 | 0.165 | - |
| Best for Bayesian | NB | IPS | FB | WPD | IPS | FB | FB | WPS | NPD | FB | NPD | FB | NPS | FB | NPS | WPS |
| | Frequentist Weibull (AFT) | | | | | | | | | | | | | | | |
| SFM.AFT | 0.007 | 0.032 | 0.032 | 107 | 0.115 | 0.037 | 0.050 | 109 | 0.655 | 0.046 | 0.475 | 108 | 0.671 | 0.034 | 0.484 | 107 |
| SIPTW.AFT | 0.012 | 0.041 | 0.041 | 93 | 0.128 | 0.039 | 0.055 | 94 | 0.433 | 0.036 | 0.223 | 96 | 0.485 | 0.031 | 0.266 | 93 |
| JFM.AFT | 0.037 | 0.060 | 0.061 | 147 | 0.058 | 0.054 | 0.057 | 147 | 0.066 | 0.054 | 0.058 | 147 | 0.074 | 0.057 | 0.063 | 148 |
| JIPTW.AFT | 0.066 | 0.057 | 0.062 | 148 | 0.061 | 0.055 | 0.059 | 148 | 0.061 | 0.056 | 0.060 | 149 | 0.073 | 0.052 | 0.057 | 149 |
| Best for AFT | SFM.AFT | SFM.AFT | SFM.AFT | SFM.AFT | JFM.AFT | SFM.AFT | SFM.AFT | JIPTW.AFT | JIPTW.AFT | SIPTW.AFT | JFM.AFT | JIPTW.AFT | JIPTW.AFT | JIPTW.AFT | JIPTW.AFT | JIPTW.AFT |
| | Frequentist (Cox) | | | | | | | | | | | | | | | |
| SFM.Cox | 0.027 | 0.029 | 0.030 | 108 | 0.324 | 0.022 | 0.127 | 108 | 0.277 | 0.032 | 0.109 | 110 | 0.392 | 0.032 | 0.186 | 109 |
| SIPTW.Cox | 0.011 | 0.034 | 0.034 | 92 | 0.254 | 0.034 | 0.098 | 91 | 0.298 | 0.033 | 0.121 | 94 | 0.396 | 0.046 | 0.202 | 91 |
| JFM.Cox | 0.043 | 0.048 | 0.050 | 146 | 0.260 | 0.040 | 0.107 | 147 | 0.235 | 0.044 | 0.099 | 147 | 0.217 | 0.049 | 0.096 | 148 |
| JIPTW.Cox | 0.038 | 0.038 | 0.040 | 148 | 0.171 | 0.057 | 0.086 | 149 | 0.143 | 0.056 | 0.076 | 149 | 0.128 | 0.070 | 0.087 | 149 |
| Best for Cox | SIPTW.Cox | SFM.Cox | SFM.Cox | JIPTW.Cox | JIPTW.Cox | SFM.Cox | JIPTW.Cox | JIPTW.Cox | JIPTW.Cox | SFM.Cox | JIPTW.Cox | JIPTW.Cox | JIPTW.Cox | SFM.Cox | JIPTW.Cox | JIPTW.Cox |
| Best from all methods | NB | IPS | FB | JIPTW.AFT | JIPTW.Cox | SFM.Cox | JIPTW.Cox | JIPTW.Cox | JFM.AFT | FB | JFM.AFT | JIPTW.Cox | JIPTW.Cox | SFM.Cox | FB | JIPTW.AFT | JIPTW.AFT |

IPD: Informative Prior Distinct; IPS: Informative Prior Same; NPD: Non-informative Prior Distinct; NPS: Non-informative Prior Same; WPD: Weakly-informative Prior Distinct; WPS: Weakly-informative Prior Same;
FB: Full-Borrowing; NB: No Borrowing; SFM: Separate Full Matching; SIPTW: Separate Inverse Probability Treatment Weighting; JFM: Joint Full Matching; JIPTW: Joint Inverse Probability Treatment Weighting